\documentclass[aps,article,prl,preprintnumbers,amsmath,amssymb,twocolumn,superscriptaddress,showpacs,floatfix]{revtex4-1}

\usepackage{epsfig}
\usepackage{dcolumn}
\usepackage{bm}
\usepackage{color}

\usepackage{amssymb}
\usepackage[figuresright]{rotating}
\usepackage{wrapfig}

\newcommand{\be}{\begin{equation}}
\newcommand{\ee}{\end{equation}}
\newcommand{\bea}{\begin{eqnarray}}
\newcommand{\eea}{\end{eqnarray}}
\newcommand{\lp}{\left(}
\newcommand{\rp}{\right)}
\newcommand{\bk}{\mathbf{k}}
\newcommand{\bq}{\mathbf{q}}
\newcommand{\bx}{\mathbf{x}}\newcommand{\bn}{\mathbf{n}}

\begin{document}

\title{Jet Flavor Tomography of Quark Gluon Plasmas at RHIC and LHC}
\author{Alessandro Buzzatti and Miklos Gyulassy}
\affiliation{Department of Physics, Columbia University, 
New York, 10027, USA}
\begin{abstract}
A new Monte Carlo model of jet quenching in nuclear collisions, CUJET1.0, is applied to predict the jet flavor dependence of the nuclear modification
factor for fragments {\it $f=\pi,D,B,e^-$} from quenched jet flavors {\it $g,u,c,b$} in central  
collisions at RHIC and LHC.
The nuclear modification factors for different flavors 
are predicted to exhibit a novel level crossing pattern over 
a  
transverse momentum range $5< p_T< 100$ GeV which can test
jet-medium dynamics
 in quark gluon plasmas.
 
\end{abstract}
\maketitle

{\em Introduction:}
Jet quenching observables provide tomographic information about the density evolution and jet-medium dynamics in quark gluon plasmas (QGP) produced in high energy nuclear collisions \cite{glv}.
Extensive studies of hard probes at the Relativistic Heavy Ion Collider (RHIC) at energies (per nucleon pair) $0.02<\sqrt{s}<0.2$ ATeV \cite{RHIC} have recently been extended to much higher c.m. energies $\sqrt{s}=2.76-5.5$ ATeV at the Large Hadron Collider (LHC).
Upgraded detectors at RHIC and the built-in heavy quark capabilities of the ALICE, ATLAS, and CMS detectors at LHC will soon open a new chapter in jet tomography by allowing for the first time the measurement of the 
jet parton flavor $a=g,u,c,b$ and mass dependence of nuclear modification factors, $R_{AA\rightarrow a\rightarrow f}(y, p_T; s ,{\cal C})$, for a wide variety of final fragments, e.g. $f=\pi,D,B,e^-$, over broader
kinematic ranges and centrality (impact parameter) classes ${\cal C}$.
We propose in this Letter that the quenched jet flavor ``spectrum'' could provide stringent new constraints on both perturbative QCD (pQCD) and string theory inspired (conformal and nonconformal) gravity dual holographic models of jet-medium dynamics in strongly interacting quark gluon plasmas.

We report results of a new Monte Carlo pQCD tomographic model \cite{CUJET}, CUJET1.0, that predicts a striking novel level crossing pattern of flavor dependent nuclear modification factors at RHIC and LHC.
This model extends the development of the GLV \cite{glv}, DGLV \cite{DGLV}, and WHDG \cite{WHDG5} opacity series approaches by including several dynamical features that require extra computational power most easily accessible via Monte Carlo techniques.
CUJET1.0 was developed as part of the ongoing DOE JET Topical Collaboration \cite{JET} effort  
to construct more powerful numerical codes necessary to reduce previous large theoretical and numerical systematic uncertainties \cite{Armesto:2011ht,Dusling:2011rz} which have hindered
quantitative jet tomography, in addition to predict new observables
that could better discriminate between dynamical models. In this
Letter, we focus on the jet flavor dependence of nuclear modification
factors.

The CUJET1.0 code features:\\
(1)
dynamical jet interaction potentials that can interpolate between the pure HTL dynamically screened magnetic \cite{DAZ} and static electric screening \cite{glv,DGLV,WHDG5} limits;\\
(2)
the ability to calculate high order opacity corrections to interpolate numerically between $N=1$ and $N=\infty$ analytic approximations;\\
(3)
full jet path proper time integration over longitudinally expanding and transverse diffuse QGP geometries;\\
(4)
the ability to evaluate systematic theoretical uncertainties such as sensitivity to formation and decoupling phases of the QGP evolution, local running coupling and screening scale variations, and other effects out of reach with analytic approximations;\\
(5)
elastic in addition to radiative fluctuating energy loss distributions;\\
(6)
convolution over $\sqrt{s}$ and flavor dependent pQCD invariant jet spectral density (without local in $p_T$ spectral index approximations);\\
(7)
convolution over final fragmentation, $D_{f/a}(x,Q)$, as well as semileptonic decay distributions.

Additional motivation for the development of the Monte Carlo based CUJET model includes addressing key open A+A phenomenology problems such as: (1) the heavy quark jet quenching puzzle discovered at RHIC \cite{RAAe05,WHDG5,MG09}, 
(2) the surprising increase of jet transparency \cite{Chen:2011vt,WHMG11,Zakharov:2011ws} of the QGP at LHC as compared to the expected linear scaling in $dN_{ch}/dy(2.76,0-5\%)=1600$ \cite{ALICEdNdy} of the jet opacity, as suggested by preliminary ALICE and CMS charged hadron quenching data \cite{ALICERAA,CMSRAA} on $R_{PbPb\rightarrow h^\pm}(p_T; \sqrt{s}=2.76,{\cal C}=0-5\%)$, and (3) the need to find hard probe observables that can better discriminate between tomographic and holographic paradigms of jet-medium interactions \cite{Horowitz:2007su}.
As we show below, the flavor dependence of the LHC nuclear modification factor level crossing pattern out to $p_T <100$ GeV could provide a rather stringent test of jet-medium dynamical models. In addition, we predict that future RHIC jet flavor tomography out to $p_T< 50$ GeV could provide important consistency checks of the pQCD paradigm due to the large $\sqrt{s}$ variation of the unquenched jet distributions between RHIC and LHC.

{\em The CUJET Model:} The CUJET1.0 code uses Monte Carlo techniques to compute finite order in opacity $N$ contributions to the jet-medium induced gluon radiative spectrum.
We replace, in the multiple collision gluon radiation kernel,
the static (Debye screened) effective potential \cite{glv,DGLV,WHDG5} [see Eq. (113) in GLV and Eq. (17) in DGLV] with a normalized but path dependent effective dynamical (magnetic enhanced) transverse momentum, ${\bf q}$, exchange distribution, generalized from the pure HTL dynamically magnetically screened model \cite{DAZ} to the form \cite{CUJET}
\be
	 \bar{v}^2(z,{\bq}; r_m)= 
\frac{{\cal N}(\mu_e(z),r_m)}{( \bq^2 + \mu_e^2(z)
             )( \bq^2 + r_m^2 {\mu}_e^2(z) )}
\;.\ee
Here $0\le r_m\equiv \mu_m/\mu_e$ is the ratio of a possible nonperturbative static color magnetic screening mass, $\mu_m\sim O(g^2T)$, to the standard static HTL color electric Debye mass, $m_e=gT$ (see also \cite{MD11}).

To illustrate the proposed jet flavor tomography test of jet-medium dynamical models, we show here results in the pure $r_m=0$ ($N=1$) HTL limit \cite{DAZ}.
Preliminary results up to third order in opacity with different $r_m$ were reported in  ref.(BG) \cite{CUJET} and do not qualitatively change the results.
In this approximation, 
the generalization to an inhomogeneous, nonstatic plasma 
of the induced radiated gluon number per collinear light cone momentum fraction $x_+$, for massive quark jets of flavor $a$ produced at position $\bx$ with transverse energy $E$ and propagating through a QGP density field $\rho_{QGP}(\bx,\tau)$ in azimuthal direction $\phi$, is given by%
\begin{eqnarray}
x_+\frac{dN_{g/a}}{dx_+}(\bx,\phi) &=&  
\kappa_a \int d\tau \rho_{QGP}(\bx+\hat{\bn}(\phi)\tau, \tau)\nonumber\\
&\;& \hspace{-0.4in} \times \int  \frac{d^2{\bq}}{\pi}
\frac{1}{\bq^2 (\bq^2{+}\mu^2(\tau))} \int \frac{d^2{\bk}
}{\pi}
  \nonumber \\
&\;& \hspace{-1.1in} \times \frac{2(\bk{+}\bq)}{(\bk{+}\bq)^2{+}\chi_a(\tau)}
	\cdot \lp \frac{(\bk{+}\bq)}{(\bk{+}\bq)^2{+}\chi_a(\tau)} - \frac{\bk}{\bk^2{+}\chi_a(\tau)} \rp \nonumber \\
&\;&   \hspace{-0.8in}\times \lp 1- \cos\left[\frac{(\bk{+}\bq)^2+\chi_a(\tau)}{2x_+E}\; \tau\right] \rp
   \;.
\end{eqnarray}
Here $\kappa_a ={9C_2(a) \alpha_s^3/2}$ with $C_2(g)=3,C_2(q)=4/3$, and $x_+$ is the fraction of plus momentum carried away by the radiated gluon.
Also $\mu(\tau)=gT(\bx+\hat{\bn}(\phi)\tau, \tau)$ is the local path dependent Debye screening mass with $\bx$ the jet production point.
$\chi_a(\tau)=M_a^2 x_+^2 + \mu^2(\tau) (1-x_+)/2$ controls the ``dead cone'' effect due to the finite jet current quark mass for $a=g,u,c,b$; ${\mu(\tau)}/{\sqrt{2}}$ is the local HTL asymptotic gluon thermal mass .
We include fluctuations of radiative energy loss 
computing $P_a(\epsilon)$, the probability distribution of radiating a fraction of energy $\epsilon$, via a Poisson convolution of the spectrum ${dN_{g}}/{dx_E} \equiv ({dx_+}/{dx_E}){dN_g}/{dx_+}$, with $x_E$ the energy fraction carried away by the radiated gluon.
As in WHDG, we convolute the Thoma-Gyulassy (TG) model for elastic energy loss assuming Gaussian fluctuations.

However, in contrast to WHDG11 and DGLV, where longitudinal expansion is taken into account via mean proper time $\tau=L/2$ approximation, the CUJET Monte Carlo integrates numerically arbitrary proper time evolution and thus allows the study of uncertainties associated with short time transient nonequilibrium QGP formation physics \cite{Dusling:2011rz} as well as variations of the jet decoupling time. 
The results shown in this Letter are obtained assuming
\be
\rho_{QGP}(\bx,\tau)= \frac{3dN_{ch}/dy}{2 N_{part}} 
\frac{\rho_{part}(\bx)}{\tau_0} \left\{ \begin{matrix}
{\tau}/{\tau_0} \;(\tau<\tau_0)\\
{\tau_0}/{\tau} \;(\tau>\tau_0)\\
\end{matrix}\right.
\ee
where $\rho_{part}({\bf x})/N_{part}$ is the normalized Glauber transverse participant nucleon density profile that depends on $A$, $\sqrt{s}$ and impact 
parameter $b$, and $dN_{ch}/dy$ is the bulk charged hadron rapidity density.
We show results for $\tau_0=0,1$ fm/c and assume a sudden thermal freeze-out hypersurface ($\rho_{QGP}=0$) with $T(\bx + {\hat{\bn}(\phi)}\tau_f,\tau_f)=T_f$ and $\mu_f=g T_f=200 MeV$ that terminates the jet path integration. The details of the level crossing pattern turn out to be surprisingly insensitive to the initial functional form of the opacity or the value of $\tau_0$ (as well as the freeze-out temperature), making such observable a robust prediction with respect to the different systematic uncertainties of the model.
The average over initial transverse jet production points is taken according to the standard binary collision Glauber $T_{AA}({\bf x},b=2\;{\rm fm})$ density profile.

For each flavor jet and initial ${\bf x},\phi$, CUJET computes the fractional energy loss, $\epsilon$, and the probability distribution $P_a(\epsilon ; p_i, \bx,\phi)$ for a specified range of initial $p_i$, including delta function contributions at $\epsilon=0$ and $1$.
CUJET then numerically averages the results over ${\bf x}$ and $\phi$ in order to obtain the final quenched partonic invariant cross section: 
\begin{eqnarray}
\label{Raa}
	\frac{d\sigma_a(p_f)}{dyd^2 p_f} &\equiv&  
 R_{AA}^a(p_f) \frac{d\sigma^0_a(p_f)}{dyd^2 p_f} \nonumber \\
&\;& \hspace{-1.0in}
=\left\langle \int 
d\epsilon\; P_a(\epsilon; p_i ,\bx,\phi )  
\left(\frac{d^2p_i}{d^2p_f}\right) 
\frac{d\sigma_a^0(p_i)}{dyd^2 p_i}\right\rangle_{\bx,\phi} 
 \;\;.
\end{eqnarray}
We have $p_i=p_f/(1-\epsilon)$.
Note that CUJET avoids the local spectral index approximation $R_{AA}^a\approx \langle (1-\epsilon_a)^{n_a(p_i)-2}\rangle $ to eliminate that possible source of numerical uncertainty. 
The pQCD initial jet flavor invariant $pp$ cross sections, $d\sigma^0_a$ for $\sqrt{s}=0.2$ and $2.76$ ATeV and $y=0$ for light gluon and quark ($a=g,u$) jet flavors, were computed from the LO pQCD CTEQ5 code of X.N. Wang \cite{XNWang} as in WHDG11 \cite{WHMG11}.
NLO and FONLL charm and bottom quark invariant cross sections for both RHIC and LHC were provided by Vogt \cite{Vogt11}. We computed $R_{AA}^{c,b}$ with both cross sections to estimate the error band associated with 
this initial jet spectral source of systematic uncertainty.

{\em RHIC and LHC Results:}
\begin{figure*}[tbh]
\label{FlavRHIC}
\includegraphics[width=3.in
]{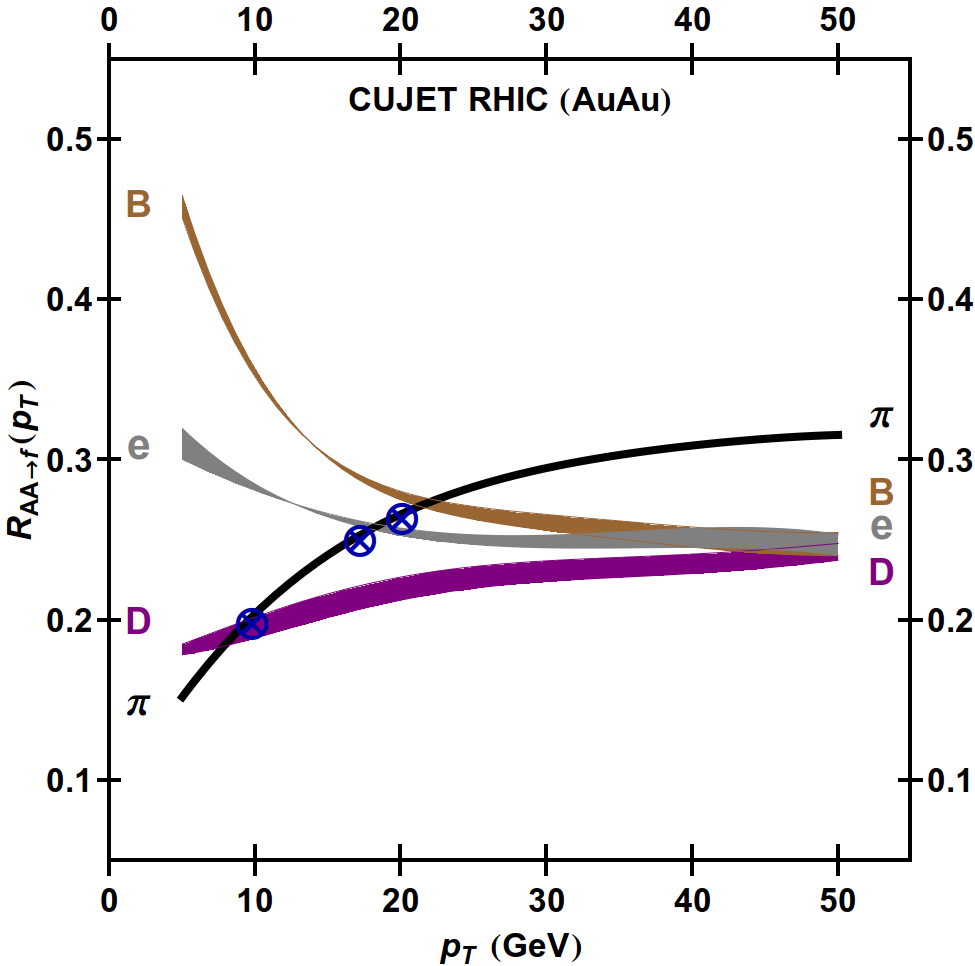}
\hspace{0.25in}
\includegraphics[width=3.in
]{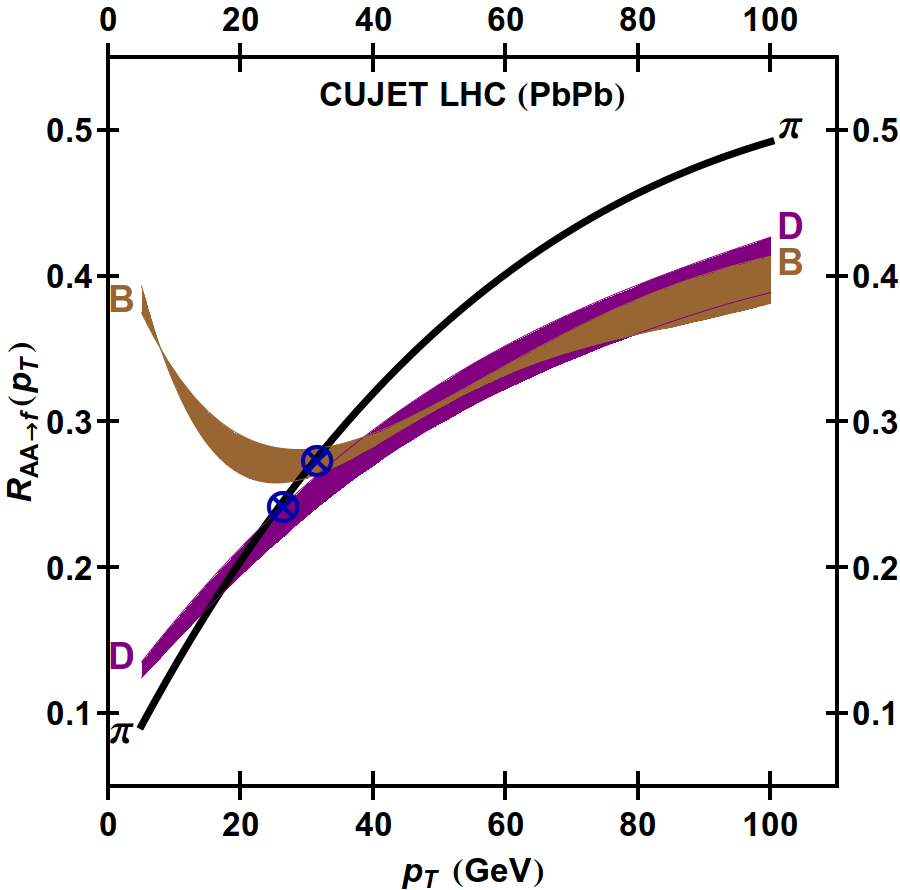}
\caption{Illustration of jet flavor tomography level crossing pattern
  of nuclear modification factors versus $p_T$ at $y=0$ for
  $\pi,D,B,e$ fragmentation from quenched $g,u,c,b$ jets in Au+Au 5\%
  at RHIC (left side) and extrapolated to Pb+Pb 5\% at LHC (right
  side) computed with the dynamic CUJET1.0 model at leading $N=1$
  order in opacity.  The opacity is constrained at RHIC, given
  $dN/dy(RHIC)=1000$ and $\tau_0=1$ fm/c, by a fit to a reference
  point $R_{AuAu}^\pi(p_T=10\;{\rm GeV})=0.2$ setting $\alpha_s=0.3$.
  The extrapolation to LHC assumes $dN_{ch}/d\eta$ scaling of the
  opacity as measured by ALICE \protect{\cite{ALICEdNdy}}.  The
  $D,B,e$ bands reflect the uncertainty due to the choice of NLO or
  FONLL initial production spectra.  Setting $\tau_0=0$ fm/c but
  readjusting $\alpha_s=0.27$ to fit our reference pion point, the (NLO)
  crossing points (crossed blue circles) are only slightly offset.
  Note the possible inversion of $\pi,D,B$ levels predicted by CUJET
  at high $p_T$ at LHC and a partial inversion at RHIC arising from
  competing dependences on the parton mass of energy loss and
  of initial pQCD spectral shapes.  }
\end{figure*}

Our central physical tomographic assumption leading to our main
result, Fig. 1, is that aside from the unavoidable $\sqrt{s}$
dependence of the initial pQCD partonic invariant cross sections, the
sole $\sqrt{s}$ dependent nuclear input in CUJET is the variation of
the bulk final pion rapidity density, $dN_{AA}/dy$.  Therefore, as in
WHDG11 \cite{WHMG11}, we assume that the charged particle
pseudo-rapidity density $dN_{ch}^{LHC}/d\eta=1600$, reported by ALICE
\cite{ALICEdNdy} for ${\cal C}=0-5\%$ central $Pb+Pb$ at $2.76$ ATeV,
translates directly into a $2.2$ increase factor of $\rho_{QGP}$ at
LHC relative to RHIC (for the same centrality).  Furthermore, given
$dN/dy=1000$, we fix the RHIC partonic level to constrain one
reference $p_T=10$ GeV point of pion $R_{AA}^\pi=0.2$ setting
$\alpha_s=0.3$ in Eq.(3). The robustness of our results, i.e. the
insensitivity of the level crossing pattern to the various systematic
uncertainties of the model, is granted by the freedom to fit the RHIC
pion reference point 
by varying the coupling parameter $\alpha_s=0.3\pm 0.03$. 
Because CUJET includes the dynamical magnetic
enhancement\cite{DAZ}, this moderate coupling
is sufficient to account for the RHIC data with $dN/dy=1000$ as we
check below (see Fig. 2).  The RHIC constrained extrapolation to LHC
is then parameter free (assuming $\alpha_s$, $\tau_0$ and $T_f$ do not
vary with $\sqrt{s}$).
A detailed study of systematic uncertainties associated with variations of the effective jet-medium coupling \cite{Zakharov:2011ws,Armesto:2011ht,Dusling:2011rz}, as well as initial and final temperature field profiles or opacity order approximation, will be reported elsewhere.

In Fig.1 on the left side, the splitting between pion and electron $R_{AA}$ is found to remain quite evident below $10$ GeV in spite of the use of the dynamically enhanced potential. Nevertheless, with respect to WHDG, the value of the electron nuclear modification factor is sensibly lower and possibly consistent with data. We believe that present uncertainties in the data do not allow to discriminate between the models.
We then show the separate contribution of D and B mesons that future detector upgrades at RHIC could test: the importance of experimentally isolating and observing charged heavy mesons cannot be overstated since the mass splitting between c and b jets is a particularly robust prediction of pQCD in a deconfined QGP medium.
The novel inversion of the $\pi<D<e<B$ $R_{AA}$ hierarchy ordering at high $p_T$ is due to the steeper initial invariant jet distributions of c and b jets at RHIC \cite{Vogt11} as also noted in \cite{Aurenche:2009dj}.
Holographic gravity dual models predict a qualitatively different heavy quark quenching pattern \cite{Horowitz:2007su}.

Future LHC data will provide an opportunity to test details of the jet flavor dynamics at densities at least a factor of two greater than at RHIC.
On the right side of Fig.1, the RHIC constrained LHC extrapolated jet flavor tomographic pattern is shown.
With the much wider kinematic window accessible at LHC, the predicted flavor dependent $p_T$ spectrum of nuclear modifications is seen to involve multiple level crossings that are qualitatively different than at RHIC energies because of the complicated interplay between flavor dependent spectral shapes and opacity enhanced jet energy loss.
While the absolute value of the nuclear modification factor depends sensitively on specific dynamical mechanics such as the effective coupling, initial formation time and freeze-out, the shape of the level crossing pattern is not and the differentials in $p_T$, $\sqrt{s}$, ${\cal C}$ and $A$ variations of the jet flavor quenching pattern appear particularly promising in order to discriminate between jet-medium dynamical models.
Of course, flavor tagged dihadron and hadron-lepton correlations will enable much more detailed quantitative information to be extracted eventually.
\begin{figure}[tbh]
\label{PiRHvsLH}
\includegraphics[width=3.in
]{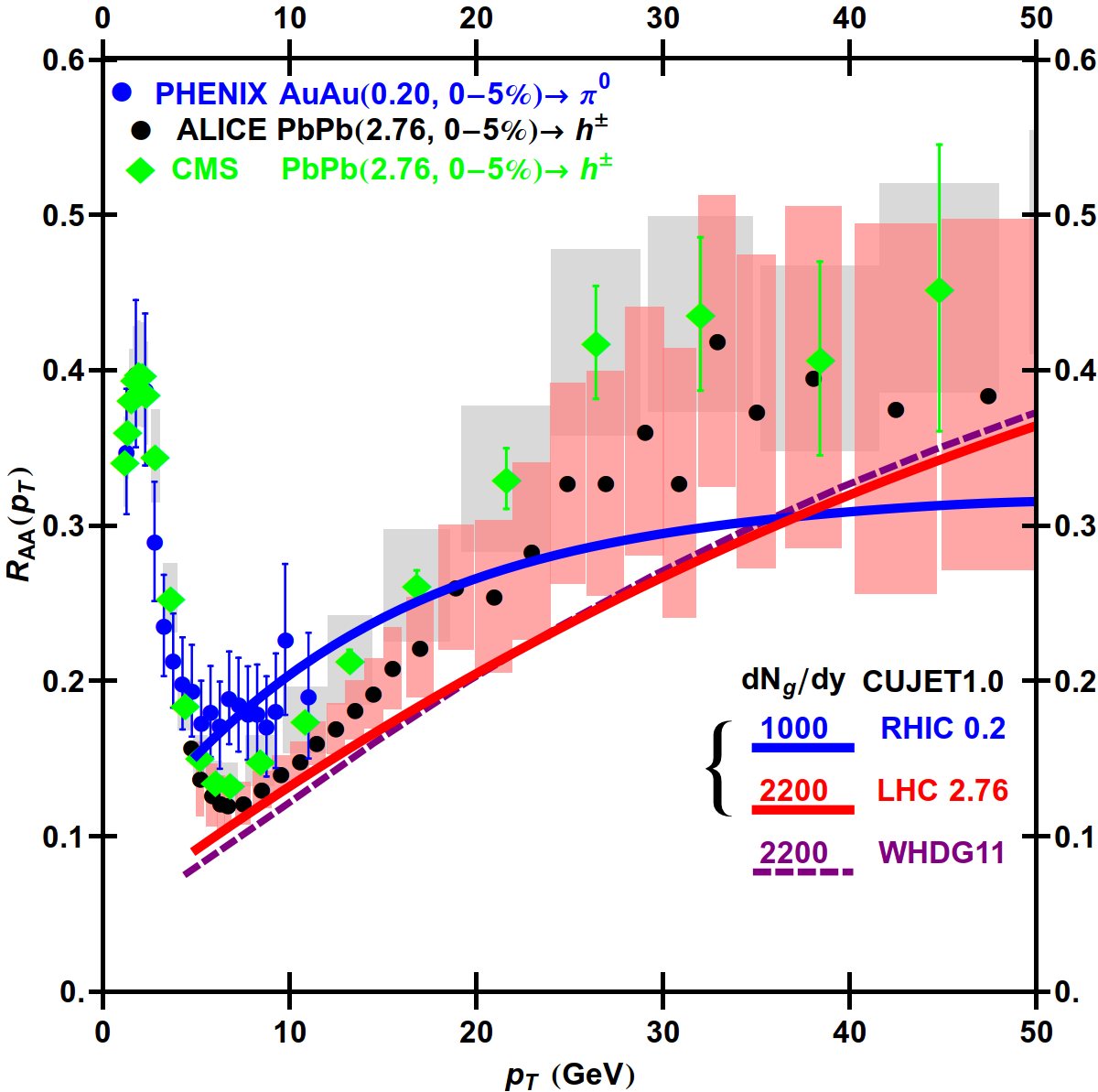}
\caption{Dynamically ($r_m=0$) enhanced jet quenching \protect{\cite{DAZ}} at RHIC (blue) and LHC (red) with the CUJET1.0 constrained 
at $p_T=10$ GeV RHIC with $\alpha_s=0.3$ as in Fig.1.
Central $0-5\%$ PHENIX $\pi^0$ RHIC data \protect{\cite{Adare:2008qa}} (blue dots) and preliminary ALICE and CMS $h\pm$ LHC data \protect{\cite{ALICERAA,CMSRAA}} (black dots and green diamonds, resp.) are compared to predictions.
For comparison, previous LHC prediction from WHDG11 \protect{\cite{WHMG11}} (purple) based on static Debye screened jet-medium interactions is shown as well.
}
\end{figure}
In Fig. 2 we compare our CUJET constrained predictions for pion $R_{AA}^\pi(p_T)$ to central PHENIX/RHIC \cite{Adare:2008qa} and preliminary ALICE/LHC \cite{ALICERAA} and CMS/LHC \cite{CMSRAA} data.
We also compare to the  LHC prediction of WHDG11 \cite{WHMG11}.
CUJET is seen to lift the nuclear modification factor at lower $p_T$ due to a complicated interplay between dynamically enhanced
but initial time reduced (for $\tau<\tau_0$) opacity evolution. 
However, both LHC curves tend to fall below the preliminary LHC data 
(as do several other recent predictions \cite{Chen:2011vt,Zakharov:2011ws}, or Debye screened potential models IV02 \cite{glv} and recently updated WHDG11 \cite{WHMG11}, not all shown here).
This suggests the intriguing possibility that the effective jet-medium coupling at LHC could be weaker than at RHIC.
See ref.\cite{WHMG11}, however, for a detailed discussion of many open caveats,
e.g. gluon feedback, $g$ to $q$ jet conversion, in this connection.
Flavor tagged single and dijet tomography can  
help to differentiate between the competing dynamical mechanisms. 

Discussions with  R. Vogt, W. Horowitz, A. Ficnar, B. Betz, J. Noronha, M. Mia,  and X.N. Wang are gratefully acknowledged. Support for this work from the US-DOE Nuclear Science Grant No. DE-FG02-93ER40764 and  No.\ DE-AC02-05CH11231 (within the framework of the JET Topical Collaboration \cite{JET}), and the CERN TH 2011 visitor program are also gratefully acknowledged.


\begin{thebibliography}{00}

\bibitem{glv} 	M. Gyulassy, I. Vitev, X. N. Wang, B. W. Zhang,
	{\em Quark Gluon Plasma 3}, (World Sci.,p.123)
	[nucl-th/0302077];    
M. Gyulassy,
  Lect. Notes Phys.  {\bf 583} (2002) 37
; 
	(IV02) I. Vitev, M. Gyulassy,
  Phys. Rev. Lett. 89, 252301 (2002); 
(GLV) M. Gyulassy, P. Levai and I. Vitev,
	Nucl. Phys. B 594, 371 (2001);
  X. N. Wang and M. Gyulassy,
  Phys. Rev. Lett.  {\bf 68} (1992) 1480.



\bibitem{RHIC}
J.~Adams {\it et al}  [STAR Collab.],
 Nucl. Phys.   {\bf 757}, 102 (2005);
  K.~Adcox {\it et al}  [PHENIX Collab.],
  Nucl. Phys. A {\bf 757}, 184 (2005).


\bibitem{CUJET}
  (BG) A.~Buzzatti and M.~Gyulassy,
  Nucl. Phys. A {\bf 855}, 307 (2011) and
to be published.
  


\bibitem{DGLV}
	(DGLV) M.~Djordjevic and M.~Gyulassy,
	Nucl. Phys. A 733, 265 (2004).

\bibitem{WHDG5}
	(WHDG) S.~Wicks, W.~Horowitz, M.~Djordjevic and M.~Gyulassy,
	Nucl. Phys. A 784, 426 (2007).


\bibitem{JET}
  (JET) Topical Collaboration on Jet and Electromagnetic Tomography,
  http://www-nsdth.lbl.gov/jet/.

\bibitem{Armesto:2011ht}
  N.~Armesto {\it et al},
  arXiv:1106.1106 [hep-ph].

\bibitem{Dusling:2011rz}
  K.~Dusling, F.~Gelis and R.~Venugopalan,
  arXiv:1106.3927 [nucl-th].



\bibitem{DAZ}
	M.~Djordjevic,
	Phys. Rev. C 80, 064909 (2009);
	M.~Djordjevic and U.~W.~Heinz,
	Phys. Rev. Lett. 101, 022302 (2008);
  P.~Aurenche, F.~Gelis and H.~Zaraket,
  JHEP 0205, 043 (2002);
  B.~G.~Zakharov,
  JETP Lett. 76, 201 (2002).
  

	
\bibitem{RAAe05}
	S.~S.~Adler {\it et al} [PHENIX Collab.]
  Phys. Rev. Lett. {\bf 96} 032301 (2006);
  B.~I.~Abelev {\it et al}, [STAR Collab.] 
  Phys. Rev. Lett.{\bf 98} 192301 (2007). 

\bibitem{MG09}
	M.~Gyulassy,
	APS Physics 2, 107 (2009).
\bibitem{Chen:2011vt}
  X.~F.~Chen {\it et al} 
  arXiv:1102.5614 [nucl-th].

\bibitem{WHMG11}
(WHDG11)
  W.~A.~Horowitz and M.~Gyulassy,
  ``The Surprising Transparency of the sQGP at LHC'',
  arXiv:1104.4958 [hep-ph].

\bibitem{Zakharov:2011ws}
  B.~G.~Zakharov,
  arXiv:1105.2028 [hep-ph].

\bibitem{ALICEdNdy}
  K.~Aamodt {\it et al}  [The ALICE Collaboration],
  Phys. Rev. Lett. {\bf 105}, 252301 (2010).


\bibitem{ALICERAA}
  K.~Aamodt {\it et al}  [ALICE Collab.],
  Phys. Lett. B {\bf 696}, 30 (2011); ALICE Collab, SQM2011 proceedings, to be published.

\bibitem{CMSRAA}
  The CMS Collab. to be published, CMS- PAS- HIN-005, 
  http://cdsweb.cern.ch/ record/1352777?ln=en

\bibitem{Horowitz:2007su}
  W.~A.~Horowitz, M.~Gyulassy,
  Phys. Lett. B 666, 320-323 (2008);
  W.~A.~Horowitz,
  Nucl. Phys. A {\bf 855}, 225 (2011);
J.~Noronha {\it et al}, 
  Phys. Rev. C 82 (2010) 054903;
  A.~Ficnar {\it et al},  
  Nucl. Phys. A {\bf 855}, 372 (2011);
  J.~Casalderrey-Solana {\it et al} 
  arXiv:1101.0618 [hep-th].


\bibitem{MD11}
  M.~Djordjevic {\em et al},
  arXiv:1105.4288, 4359 [nucl-th].


  



\bibitem{XNWang} X.N.Wang, Private communication.
\bibitem{Vogt11}
  M.~Cacciari, P.~Nason and R.~Vogt,
  Phys.\ Rev.\ Lett.\  {\bf 95}, 122001 (2005); R.Vogt Private communication.



\bibitem{Aurenche:2009dj}
  P.~Aurenche and B.~G.~Zakharov,
  JETP Lett.\  {\bf 90}, 237 (2009).

\bibitem{Adare:2008qa}
  A.~Adare {\it et al.}  
  Phys.\ Rev.\ Lett.\  {\bf 101}, 232301 (2008).




\end{thebibliography}
\end{document}